\documentclass[a4paper, table]{article}

\usepackage[utf8]{inputenc} 
\usepackage{amsmath, amsthm, amssymb, amsfonts} 
\usepackage{lipsum} 

\usepackage{graphicx} 
\usepackage{float} 
\usepackage{caption} 
\usepackage{subcaption} 

\usepackage{tabularx} 
\usepackage{tabu} 
\usepackage{booktabs} 

\usepackage[nottoc,numbib]{tocbibind} 
\usepackage[margin = 2.5cm]{geometry} 
\usepackage{microtype} 
\usepackage{titlesec} 
\usepackage{titletoc} 
\usepackage{appendix} 
\usepackage{fancyhdr} 
\usepackage[shortlabels]{enumitem} 

\usepackage{hyperref} 
\usepackage[noabbrev, capitalise]{cleveref} 

\usepackage{xcolor} 

\usepackage{tikz} 
\usetikzlibrary{positioning} 
\usetikzlibrary{calc} 

\makeatletter
\newcommand{\gettikzxy}[3]{%
  \tikz@scan@one@point\pgfutil@firstofone#1\relax
  \edef#2{\the\pgf@x}%
  \edef#3{\the\pgf@y}%
}
\makeatother

\newcommand\reporttitle{\baselineskip=32pt Closure models for the feedback of \\  \bigskip energetic particles on plasma turbulence}

\newcommand\reportsubtitle{ 
}

\newcommand\groupnumber{
\textbf{}
}

\usepackage{orcidlink}
\newcommand\reportauthors{
 Jane Pratt \orcidlink{0000-0003-2707-3616} (pratt34@llnl.gov)
 \\
 (Georgia State University, Atlanta, USA, and
 \\
 Physics Division, Physics and Life Sciences, LLNL)

}

\newcommand\grouptutor{
}

\newcommand\placeanddate{
Livermore, California \today
}

\definecolor{Tue-red}{RGB}{199, 25, 24}
\definecolor{lightblue}{rgb}{.8, 1., 1.}
\definecolor{cadet}{rgb}{.3725, .619, .627}
\definecolor{cyan}{rgb}{0., .545, .545}
\definecolor{sea}{rgb}{.235, .702, .443}
\definecolor{aqua}{rgb}{.561, .737, .561}
\definecolor{turq}{rgb}{.686, .9333, .9333}
\definecolor{whiteblue}{rgb}{0.2, .8, .6}
\definecolor{bluey}{rgb}{0.2, .8, 1.}
\definecolor{ltblue4}{rgb}{.902, .957, 1}
\definecolor{seablue}{rgb}{.3725,  .619,  .627}
\definecolor{dodger}{rgb}{.0,.2758,.5151}

\hypersetup{
    colorlinks=true,
    linkcolor=dodger,
    urlcolor=dodger,
    citecolor = dodger
    }
\counterwithin{equation}{section} 
\counterwithin{figure}{section} 
\counterwithin{table}{section} 

\titleformat{\section}{\sffamily\color{dodger}\Large\bfseries}{\thesection\enskip\color{gray}\textbar\enskip}{0cm}{} 

\titleformat{\subsection}{\sffamily\color{dodger}\large\bfseries}{\thesubsection\enskip\color{gray}\textbar\enskip}{0cm}{} 

\titleformat{\subsubsection}{\sffamily\color{dodger}\bfseries}{\thesubsubsection\enskip\color{gray}\textbar\enskip}{0cm}{} 

\DeclareCaptionFont{dodger}{\color{dodger}}
\usepackage{mlmodern}

\usepackage{graphicx}
\usepackage{xspace}
\usepackage{amsmath}
\usepackage{amsfonts}
\usepackage{amssymb}
\usepackage{placeins}
\usepackage{gensymb}
\usepackage{color}
\usepackage{float}
\usepackage{hyperref}
\usepackage[english]{babel}
\usepackage{natbib}
\usepackage{ragged2e}
\usepackage[many]{tcolorbox}    

\definecolor{orange}{rgb}{.95,.3,0}

\newtcolorbox{abst}{enhanced, colback=gray!60, colframe=gray!60, width=\linewidth,size=fbox, oversize=2pt, boxrule=0pt, toprule=1pt, bottomrule=1pt, sharp corners, boxsep=3pt, top=6pt, bottom=6pt, breakable=true,coltext=black,fontupper=\sffamily\upshape, title={{\color{black}ABSTRACT}}}

\begin{document}

\begin{titlepage}

\centering

\begin{tikzpicture}

\node[opacity=0.2,inner sep=0pt,remember picture,overlay] at (4.5,-0.5){\includegraphics[width= 0.8 \textwidth]{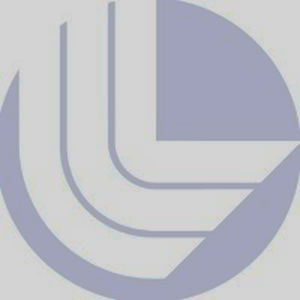}};

\node[inner sep=0pt] (logo) at (0,0)
    {\includegraphics[width=.25\textwidth]{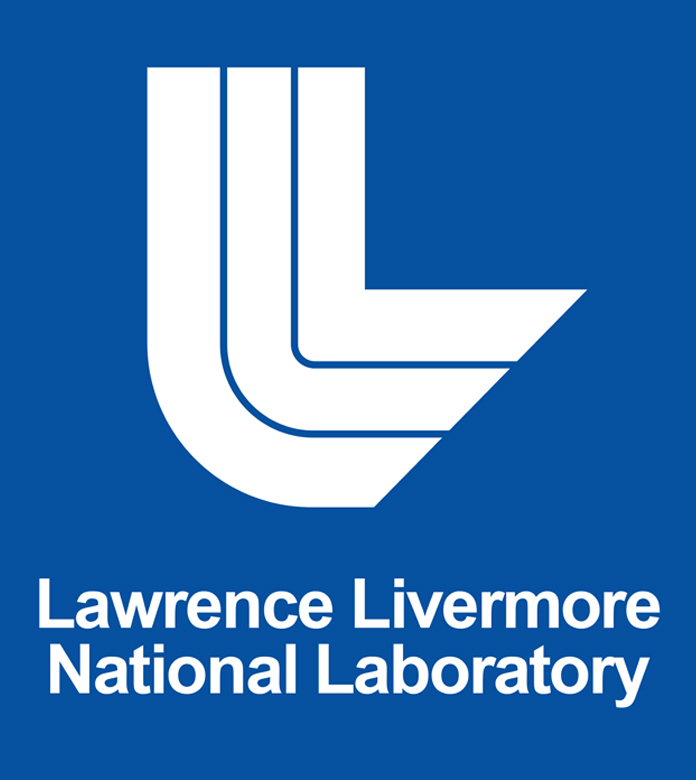}};
    
\node[text width = 0.7\textwidth, right = of logo](title){\sffamily\huge\reporttitle};

\node[text width = 0.5\textwidth, yshift = 0.75cm, below = of title](subtitle){\sffamily\Large \reportsubtitle};

\gettikzxy{(subtitle.south)}{\sffamily\subtitlex}{\subtitley}
\gettikzxy{(title.north)}{\titlex}{\titley}
\draw[line width=1mm, dodger]($(logo.east)!0.5!(title.west)$) +(0,\subtitley) -- +(0,\titley);

\end{tikzpicture}
\vspace{3cm}

\sffamily\groupnumber

\begin{table}[H]
\centering
\sffamily
\large
\begin{tabu} to 0.8\linewidth {cc}

\sffamily\reportauthors

\end{tabu}

\end{table}

\sffamily \grouptutor

\tikz[remember picture,overlay]\node[anchor=south,inner sep=0pt] at (current page.south) {\includegraphics[width=\paperwidth]{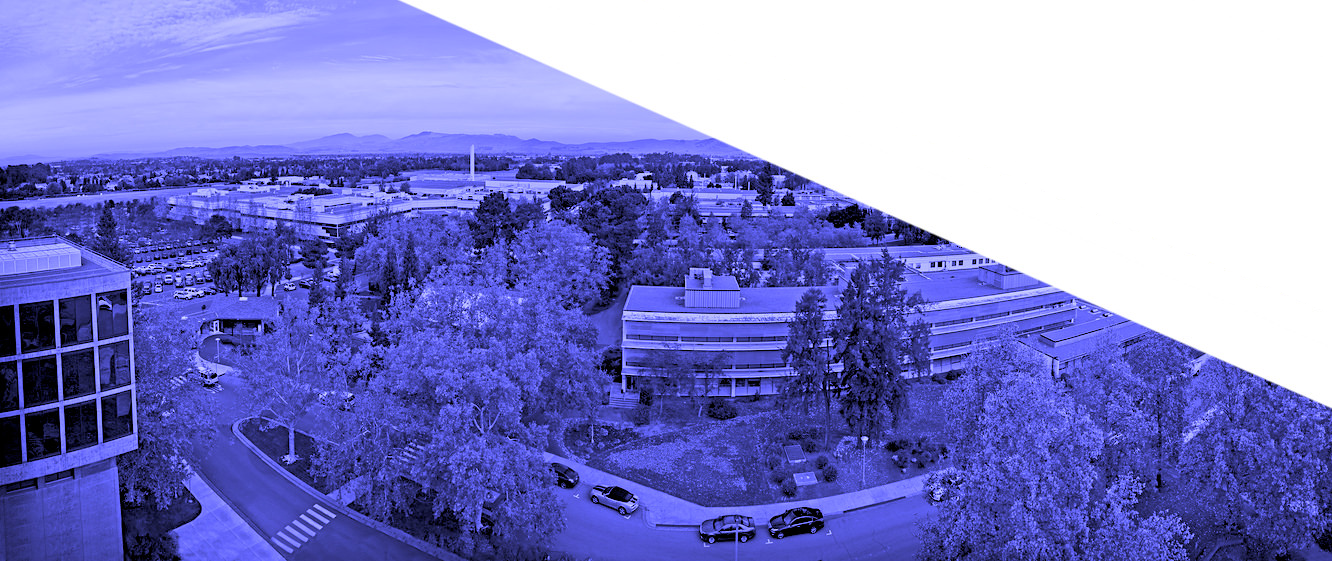}};

\mbox{}
\vfill
\sffamily \Large \textcolor{white}{\placeanddate} \\

\end{titlepage}

\newpage

\baselineskip=8pt
{\hypersetup{linkcolor=black} 
\tableofcontents\thispagestyle{empty}}

\vspace*{10mm}
\normalsize
\justifying{ 
\begin{abst}
Energetic particles interact with the plasma surrounding them, resonating with certain types of plasma waves to stabilize them while destabilizing others, and changing the character of the background turbulence in ways that have not been fully quantified or understood.   Interaction with the turbulent background plasma is key to the acceleration of many types of energetic particles including high-energy cosmic rays,  solar energetic particles, and pick-up ions. The acceleration of particles is a process that would ideally be described by a kinetic model, a type of model that follows a probability distribution function (PDF) for all particles in {7-dimensional} $(x, y, z, v_x, v_y, v_z, t)$ space.  Because of the high dimensionality of a kinetic model, simulations that solve kinetic equations use the largest computational resources currently available, and are yet unable to simulate a realistic number of particles, reach the large scales necessary for astrophysical problems, and use high-precision numerical methods.   Two available alternatives to kinetic plasma models have been explored for this problem, with limited success.  One is a multi-fluid model produced by a cumulant discarding closure, which evolves coupled equations for the velocity, magnetic field, and internal energy for both the background plasma and the fluid of energetic particles.  However, simulations that solve multi-fluid magnetohydrodynamic (MHD) equations are able to include the interaction with energetic particles only in crude ways, typically as an {add-on} pressure term.  The second alternative is to use a hybrid method to couple a fluid description of the background plasma to a kinetic model or a Fokker--Planck model for the energetic particles.  These methods are hampered by the physical modeling of the coupling.  In this work, we develop a new model, which follows the PDF for all particles; this can be viewed as a step toward physical realism above a multi-fluid MHD model, while also being more computationally efficient than a kinetic model.  The equations we develop model both the background plasma and the energetic particles self-consistently.  Over the last decade, similar PDF methods have been developed to a high level of sophistication to model reactive flows and turbulent combustion for engineering applications.    For treatment of the feedback of the energetic particles on a background plasma, a PDF closure approach should evaluate the mean characteristics, including the density, with better statistical quality than will particle-sampling procedures. 
\end{abst}
}
 \pagenumbering{arabic}
 \baselineskip=14pt


\section{\label{sec:intro} Introduction}

Astrophysical research into energetic particles began with the discovery of cosmic rays, a rich field that spans the entire 20th century.  Cosmic rays are extremely high-energy particles understood to be accelerated by interaction with astrophysical plasmas.  How the hottest cosmic rays, the ultra high energy cosmic rays, become so much hotter and faster than other energetic particles is a central question.  Theories for the acceleration of these particles center on their interactions with plasma shocks.  One situation where this occurs is a supernova explosion \citep{bz1934,lagage1983maximum,axford1992particle,uchiyama2007extremely,bykov2012galactic}.  During a supernova explosion, particles are ejected into the interstellar medium, and some of those particles are thought to interact intensely with the expanding shock-wave.  There are many other places in the universe where plasma shocks occur, including the shock surrounding the Local Bubble, the solar wind termination shock, interplanetary traveling shocks, accretion shocks, the heliospheric shock, and Earth's bow shock.  In any of these settings, particles could be accelerated by interacting with the shock. 

The theory for acceleration of energetic particles near shocks is unfinished.  In practice, it is likely that multiple plasma processes contribute to the remarkably efficient acceleration that leads to high-energy cosmic-rays.   The most frequently studied mechanisms for cosmic-ray-acceleration are diffusive shock acceleration \citep[e.g.][]{drury1983introduction, berezhko1999simple, malkov2001nonlinear,caprioli2024particle}, and stochastic acceleration \citep[e.g.][]{petrosian2012stochastic,hartquist1983evidence,walter2025stochastic,globus2025ultrahigh}.   Considering hybrid acceleration processes, for example, particles could be initially accelerated by interaction with turbulence (stochastic acceleration) and subsequently accelerated further by interacting with a shock.  Magnetic reconnection can also accelerate particles, particularly when magnetic fields are compressed or strongly varying \citep{sironi2014relativistic, giacalone2012,somov2013particle,sakai1988particle}; this could also play a role in a combined acceleration process.

When a particle drifts across a shock-front from upstream to downstream, it gains a small amount of energy by means of a process called shock-drift acceleration (SDA) \citep[see also][]{chapter11particleaccel}.   Although the component of the magnetic field normal to the shock-front is continuous, the tangential component of the magnetic field may change abruptly.   The total magnitude of the magnetic field thus changes on a shock front.
A particle gains perpendicular velocity as it drifts across an approaching shock because the magnetic moment, $\mu=v_{\perp}^2/|B|$, is conserved to a good accuracy \citep{whipple1986adiabatic}.    This simple picture of SDA does not take into account inhomogeneities in the magnetic field, which can eventually lead to higher energy gain for the particle.

\begin{figure}
\begin{center}
\resizebox{3.8in}{!}{\includegraphics{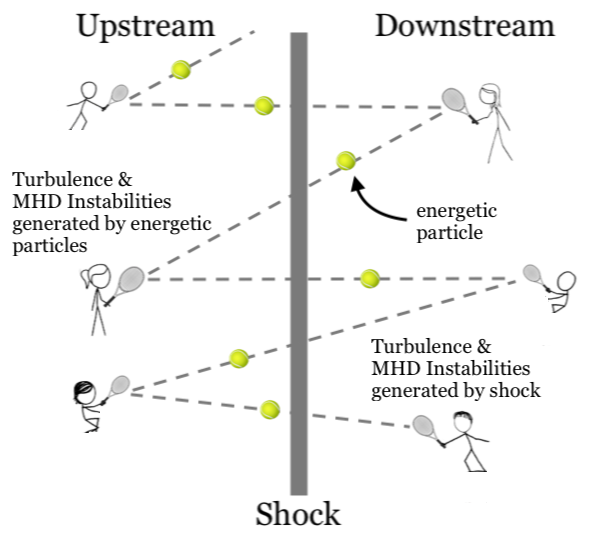}}
\caption{An illustration of the diffusive shock acceleration process by which particles gain large energies.   Tennis players represent the turbulence and instabilities, generated in different ways.  They repeatedly reflect particles (tennis balls) across the shock (tennis net). Tennis players are drawn in the style of the xkcd webcomic {(}\url{https://xkcd.com/}{).}
\label{dsacartoonfig}}
\end{center}
\end{figure}

In realistic space plasmas, the magnetic field is inhomogeneous and shocks typically propagate through a turbulent plasma with a high Reynolds number.  Magnetic turbulence and magnetic disturbances are generated downstream of a shock by the shock itself, and upstream of the shock by energetic particles \citep{bell1978a,bell1978b}.
Because the particles scatter off of magnetic inhomogeneities, particle momentum is randomized near the shock and the velocity distribution becomes isotropic in the frame of the magnetic disturbances.
When particles diffuse across the shock in either direction, they encounter a strongly turbulent plasma that (in their rest-frame) moves towards the shock-front. 
Eventually particles can diffuse across the shock-front multiple times, gaining the same small boost in energy with each crossing.  This process is called diffusive shock acceleration (DSA) \citep{longair1994high}, or first-order Fermi acceleration,
and it yields a momentum gain proportional to the velocity ratio between the shock and the particle.   Figure~\ref{dsacartoonfig} provides a cartoon of how DSA works.

For particles to reach high energies through DSA, they must remain close to the shock wave for a long time; one theoretical idea is that this could be aided by a strong magnetic field.  Since high-energy particles diffuse parallel and perpendicular to a strong magnetic field in characteristically different ways \citep[e.g.][]{beresnyak2011numerical,shalchi2004nonlinear}, a strong magnetic field  further complicates this picture of what happens near the shock.  A delicate balance is also required, because if many particles are confined near a shock for a long time, the particle pressure could smooth out the shock.  The structure of the shock and the regions around it are determined by the characteristics of  the particles that interact with the shock \citep{le1997influence}.   Active research includes how DSA works for particles moving at different angles to the shock, how different types of particles are accelerated by moving shocks \citep{yang2011acceleration,yang2012impact,chapman2005perpendicular}, and the effect of collisions \citep{malkov2012mechanism}.  
Over decades researchers have sought to define the complex cases where the shock and magnetic disturbances propagate at relativistic speeds \citep[e.g.][]{kirk1999particle,kirk2000particle,achterberg2001particle,niemiec2006cosmic2, lemoine2012particle,perri2022recent}.

Although DSA efficiently accelerates particles only when strong turbulence is present, a turbulent background plasma can also accelerate particles in a more general situation.
Stochastic acceleration, also called second-order Fermi acceleration,  functions by wave-particle interaction, as particles resonate with parts of the wave spectrum \citep{giacalone2012}.
Fermi's original argument centers on the numerous random collisions that a particle has within a magnetic cloud, which can lead to a relative momentum
gain proportional to the velocity ratio between the shock and the particle to the second power.  Stochastic acceleration is thought to play a significant role in magnetic reconnection regions and turbulent settings like molecular clouds \citep{dogiel2005cosmic}.  
Although it is not as efficient as DSA, stochastic acceleration is a sufficiently powerful mechanism to account completely for the acceleration of populations of particles to moderately high energies  \citep[e.g. as discussed by][]{lee2012shock}.    

A two-way interaction between energetic particles and turbulence is thus expected: the feedback of particles on the plasma turbulence during stochastic acceleration can lead to damping \citep{selk2007}, while the strongly nonlinear turbulence around a shock changes the characteristics of particle movement \citep{chandran2000scattering,yan2002scattering}.  Studies have shown that cosmic rays can change the character of turbulence near a shock,  generating vorticity, causing magnetic stretching and changing the magnetic topology to enhance the magnetic field \citep{zankpickup2000,marcowith2010postshock,parizot2006observational,lazarian2012turbulence}.   If cosmic-ray feedback on the turbulence results in magnetic field amplification, the altered magnetic field can make the stochastic acceleration process more efficient \citep{schure2012,downes2014cosmic}.   Nevertheless the interaction of energetic particles with the turbulent plasma is subtle, and not well understood.  The forefront of research in this area studies how stochastic acceleration is affected when turbulent scales are compressed at the shock front \citep{marcowith2010postshock},  when the turbulence is dominated by viscosity, or when the turbulence is imbalanced, i.e. when unequal energies of Alfv\'en fluctuations propagate parallel and anti-parallel to the background magnetic field because cross helicity is present \citep{lazarian2012turbulence}.

The ways that energetic particles excite and modify magnetic instabilities in the background plasma is even more complex than the mechanisms for particle acceleration.  Energetic particles are hotter, less dense, and more buoyant than the background plasma with which they interact.  They can exert pressures comparable to thermal, radiation, magnetic, and turbulent pressures.   Because of this, energetic particles have been implicated in the evolution of a great variety of magnetic disturbances.  A short list includes:\begin{enumerate}
\item  Alfv\'en waves \citep{zira2008,bell2012}.
\item Weibel instabilities \citep{weibel1959spontaneously,nishikawa2007particle}.
\item Magnetic filamentation instabilities \citep{reville2012filamentation}.
\item Gyroresonant instabilities \citep{yan2008cosmic}.
\item Firehose instabilities \citep{scott2017cosmic}.
\item Current-driven instabilities \citep{riquelme2009nonlinear,OhiraKirk2009}.
\item Non-resonant hybrid instabilities \citep{matthews2017amplification,caprioli2018diffusive,bell2004turb,reville2007cosmic}~.
\end{enumerate}
Hydromagnetic waves excited around a shock by cosmic rays are thought to contribute to cosmic-ray acceleration by amplifying the magnetic field, enhancing the turbulence, or scattering the particles \citep{reville2007cosmic,RevilleKirk2008, MalkovDiamond2006}.
While fusion scientists study many of these instabilities in laboratory plasmas, in the cosmic ray setting, the length and time scales are extreme \citep{bell2012cosmic,ryu2003cosmological}. Large-scale processes like the galactic winds and the galactic dynamo are driven by cosmic-ray streaming \citep{uhlig2012galactic,siejkowski2010cosmic,ruszkowski2023cosmic,thomas2025thermally}.  Cosmic rays may play a role in star formation \citep{mckee2007theory,murray2010disruption,papadopoulos2011extreme} and energetic particles are a dangerous component of space weather and solar storms causing damage to infrastructure on earth \citep{kataoka2012anomalous,langner2006effects,kudela2000cosmic}.

\section{Existing models for energetic particle acceleration \label{sec:existingmodels}}

Several methods have been used in simulation efforts for energetic particles, including
 kinetic models,
 multi-fluid models,
 and hybrid models in which the background plasma is treated as a magnetohydrodynamic (MHD) fluid while the energetic particles are treated either kinetically or with a Fokker--Planck model.
Here we briefly survey how these types of approaches have been used to model energetic particles.  We discuss what problems can arise with each method, and mention a few recent pieces of work that have been performed with each type of code, providing perspective and comparison between different simulation methods.

For fidelity of particle movement, the most attractive of these approaches is a kinetic model \citep[e.g. as discussed by][]{drury1983introduction}.  Several  groups have worked on cosmic-ray shock-acceleration with kinetic models, which they solve using particle-in-cell (PIC) methods, including \citep{bai2015magnetohydrodynamic,sironispit2011}, \citep{OhiraKirk2009}, and \citep{niemiec2008production}.  Many of these works  involve simulations of cosmic rays with a pre-defined background plasma and shock front. Three-dimensional kinetic/PIC simulations have been performed, but most studies have been two-dimensional due to the extraordinary computational resources that a kinetic/PIC code requires.  

Kinetic models solve for the evolving probability density function (PDF) of particles in a {7-dimensional} space consisting of spatial variables, velocity (or momentum) variables, and time.  A kinetic treatment is highly desirable because it {describes} the movement of physical particles, but this is also computationally intensive; solving a kinetic equation using a PIC method makes computation with these models accessible.  PIC methods, a numerical method for solving partial differential equations, were developed at Los Alamos in the 1950s \citep{harlow1988pic,harlow1955laur} as an efficient solver for the most demanding fluid problems in an era when computing power was more limited than today.  In a PIC formulation, phase-space is sampled at discrete points which are called ``particles''.
The simulation volume is initially prepared with these particles, and an initial equilibrium electromagnetic field is set.
  The basic equations of motion are integrated to determine the movement of PIC particles through the next time-step.  From the resulting positions and velocities of the PIC particles, current and density fields are calculated and used in discretized forms of the equations of motion to obtain magnetic and electric fields. The fields are then used to determine the movement of the PIC particles through the next time-step.   
  
While kinetic models are desirable, PIC codes have well documented limitations. When a PIC method is used, particle collisions usually need to be treated with
a Monte Carlo algorithm, and in this additional modeling physical realism is lost.  In PIC simulations that study energetic particles moving around a shock, many PIC particles eventually move downstream in the flow and leave the simulation volume. More PIC particles must be injected to maintain a sizable, continuous current and density of energetic particles.  The details of how these particles are injected is delicate, can result in multiple solutions, and can affect the quality of the numerical solutions obtained \citep[e.g. as discussed by][]{blasi2005role}.   The most serious challenge for PIC simulations is that obtaining high-quality statistics is difficult because of the limited number of PIC particles used.
The largest scales of astrophysical systems are not accessible to PIC simulations because of computational limitations.

The Fokker--Planck equation represents a simpler alternative to kinetic equations like the Boltzmann equation  \citep[e.g.][]{shalchi2009nonlinear,schlickeiser2011new,schlickeiser2013cosmic,zweibel2017basis,lasuik2017solutions,malkov2018propagating}.  The Fokker--Planck equation has the general advantage that the equation no longer needs to be solved in {7-dimensional} space, but depends only on 3 spatial variables and time (some Fokker--Planck equations still use a {7-dimensional} space).
A Fokker--Planck equation requires the calculation or modeling of additional terms: the drift and diffusion coefficients \citep[e.g.][]{schlickeiser2011new}.   The accuracy of a simple diffusion approximation for cosmic ray transport is limited, and more accurate diffusion models such as a telegraph equation  \citep{litvinenko2013telegraph,malkov2015cosmic,tautz2016application,rodrigues2019fickian}, or simply hyperdiffusion \citep{litvinenko2016comparison} have been proposed.  Cosmic-ray studies based on the Fokker--Planck equation often fall under the umbrella of hybrid methods because they are coupled to fluid equations.  In this work, our goal is to produce a model that is better tailored to the problem of energetic particles around shocks and MHD turbulence than the Fokker--Planck equation.

Numerical work on the movement and acceleration of test-particles around shocks has been performed with Monte-Carlo simulations.
Monte Carlo studies of energetic particles tend to focus on special conditions where the Fokker--Planck equation is difficult to solve, e.g. when MHD waves around the shock have extremely high amplitudes.
 Monte Carlo techniques are based on determining what fraction of randomly-generated numbers obey some requirement. 
  The basic Monte Carlo approach to energetic particles is to assume that particles move in a regular magnetic-field along undisturbed trajectories.  Scattering of the Monte Carlo particles due to particle collisions or turbulence is modeled using discrete, uncorrelated, small-amplitude perturbations to the angle between the particle's velocity vector and the local magnetic-field.  The mean time between these scattering acts and the scattering amplitudes are input-parameters used to model the magnetic fields and turbulence.  This approach to scattering is called a \emph{pitch-angle diffusion model} \citep{kirk1987particle, Ostrowski199b}, and provides a rough general approximation to homogeneous isotropic MHD turbulence.  The results from this type of Monte Carlo simulation depend on the particle injection rate and the scattering model adopted \citep{wangwang}.  A twist on this Monte Carlo method specifies the magnetic field near the shock, and integrates the equations of motion for the energetic particles in that magnetic field \citep{niemiec2006cosmic1, Lemoine2003}.
  
Because they {straightforwardly} model particle collisions and scattering, the Monte Carlo method has often been used to study energetic particles.  The Monte Carlo technique solves linear-operator equations statistically, exploiting random number generation.  While Monte Carlo methods can be extremely powerful, they require attention to statistical precision and are usually computationally intensive.        
 Recent Monte Carlo studies of cosmic rays around shocks include \citet{Lemoine2003,lemoine2006relativistic,Marcowith2006}, \citet{niemiec2006cosmic1,niemiec2004}, and many others \citep{kang1997diffusive,Kachelriess2007mc,Blasi2004mc,hiro2003mc,sigl2003mc,wolff2015cosmic}.
  
 
Two large Monte Carlo simulation efforts used to simulate energetic particle transport in tokamaks are the ASCOT and HAGIS codes \citep[see][]{peybernes2025gpu,akaslompolo2016synthetic,graves2012control,aho2008ascot,tholerus2012monte,lauber2007ligka,pinches1998hagis}.  The ASCOT code is a guiding-center particle-orbit-following Monte Carlo code that treats particles in five-dimensional phase space (three spatial and two velocity coordinates).  Simulations produced with ASCOT typically ignore turbulence, and use a fixed magnetic field for the background plasma.  ASCOT has been used, for example, to simulate sharp structures in the ion deposition on the divertor \citep{kurki2002monte}.
The HAGIS code uses a $\delta f$ Monte Carlo method.  In this formulation, the perturbed magnetic equilibrium is computed from nonlinear drift-kinetic guiding-center equations.  Energetic particle movement and collisions are modeled with the standard Monte Carlo pitch-angle diffusion model \citep[see also][]{poli2003monte}. HAGIS has been widely used for the study of energetic particle interaction with MHD instabilities.




A multi-fluid code {solves coupled sets of compressible MHD equations} for a quasi-neutral background plasma and for energetic species \citep[see][for early reviews of the multi-fluid technique]{malkov2001nonlinear,jones1991plasma}; each fluid has its own velocity field.  The basic limitation of this type of simulation is the level of sophistication of the fluid closure adopted \citep{drurytalk2009}.  
Using a fluid closure, the compressible MHD equations are coupled to energetic particles with a feedback term.  We provide a brief overview of three efforts -- {led} by the PIERNIK group, the Chandran group, and Avinash and Zank respectively --  that explore different aspects of the feedback from the cosmic rays on the background plasma through multi-fluid simulations.  While this is not an exhaustive list of work done on this problem, the work from these three groups is prominent and illustrates well the approach to, and challenges of, this problem.




The PIERNIK code \citep{zeus2010,zeus2012a,zeus2012b,hanasz2009cosmic,peschken2023phase,ogrodnik2021implementation,baldacchino2025implementation} performs two-fluid MHD simulations on an even grid over a simulation volume far larger than a supernova.  PIERNIK has been used to study the interstellar-medium dynamo driven by cosmic-ray buoyancy, and the effect of dust on the streaming instability in a proto-planetary disk \citep{kowalik2013streaming}.    PIERNIK solves the compressible MHD equations for the background plasma and, for the cosmic rays, additional equations for the energy of the cosmic-ray fluid and a cosmic-ray equation of state.  The closure model they use includes external gravity as source term, self-gravity of fluids, and a current-dependent resistivity.   The influence of the magnetic field is neglected in the plasma energy equation.  Feedback from the cosmic-ray fluid on the background plasma acts through a simple cosmic-ray pressure term in the momentum equation for the background plasma.  

The group of \citet{chandran2008convection} has studied buoyancy instabilities in galaxy clusters, using a more complex closure for the compressible MHD equations that includes both buoyancy and thermal effects.  Still more complex multi-fluid simulations have been performed by the group of \citet{avinash,avinash2007magnetic} to study magnetic instabilities in the heliosheath.  These simulations use a closure of the fully-nonlinear multi-fluid compressible MHD equations that includes charge-exchange and Hall MHD terms, and separate species for interstellar neutral particles, pick-up ions, solar-wind energetic particles, energetic electrons, and a background plasma.  The equations for these multi-fluid closures thus become significantly complicated as additional targeted effects are {added on}.

The most popular approach toward modeling cosmic rays has been hybrid simulation, where the background plasma is evolved by solving the equations of the magnetohydrodynamic (MHD) fluid approximation \citep{chandrasekhar:book} and energetic particles are treated with a kinetic-PIC model or a Fokker--Planck solver.  Hybrid codes can be differentiated by the level of sophistication of the particle feedback on the background plasma.  Some codes use fixed electromagnetic fields, calculated from a separate MHD simulation for the background plasma, and then solve for small perturbations to the magnetic and electric fields based on the contributions of energetic particles.  In the majority of cases, however, no feedback on the plasma from the energetic particles is considered.  Here we briefly survey several representative studies that use hybrid methods for modeling energetic particles.

 \citet*{reville2012filamentation} couple a kinetic model of cosmic rays to a three-dimensional incompressible ideal MHD model of the background plasma; in this work feedback on the plasma happens through a cosmic-ray current term.  \citet{beresnyak2011numerical} use an incompressible ideal MHD code to produce magnetic and electric fields, and then trace particles to understand turbulent diffusion.  Both of these groups are at the forefront of cosmic ray research, but these works neglect the feedback on the turbulence.    The MHD model that these groups use, incompressible ideal MHD, has limited applicability to cosmic ray science because this model excludes resistive phenomena like reconnection and compressible phenomena like shocks.    Later studies like that of \citet{bai2015magnetohydrodynamic} have used a similar hybrid model based on compressible ideal MHD; however, this still excludes resistive phenomena.

\citet{bell2013cosmic} investigate the escape of cosmic-ray particles upstream of the shock and how this is affected by magnetic field amplification.  They 
use two and three-dimensional MHD+kinetic simulations for cosmic ray acceleration that include feedback on the background plasma through a cosmic-ray current term.  In this study, they neglect the role of pressure gradients and magnetic tension on the feedback mechanism.

The CRASH code \citep{jones2005,miniati2001,kangryu2011,kang2007self,kang2012nonthermal} has been used to study DSA around specific types of shocks, shock self-similarity, and the energy spectra of cosmic rays.  The code treats energetic protons and electrons separately with kinetic equations.  
In CRASH simulations, the form for the distribution function in the kinetic equations is derived from linear theory.  This limits realistic CRASH simulations to studies of highly supersonic shocks and shocks strongly modified by the cosmic ray pressure.  The CRASH code uses the kinetic equations to solve for a pressure due to the cosmic rays, which feeds back on the background plasma through the momentum equation.  This pressure calculation is numerically involved because the pressure must be calculated by integrating to obtain the correct moment of the kinetic equation.   In these works, the CRASH code also does not evolve the magnetic field, nor do magnetic-field terms from the background plasma contribute to the momentum equation or pressure equation of state. 

In a similar spirit to these hybrid fluid+kinetic approaches, \citet{boss2023crescendo} have included a Fokker--Planck solver for the cosmic-rays in a variant of the GADGET code, which solves the smooth particle hydrodynamics (SPH) equations.  \citet{pfrommer2017simulating} have coupled a Fokker--Planck solver for the cosmic-rays which contributes to a pressure in the fluid moving mesh code AREPO.  Both of these modeling efforts use the cosmic ray pressure to couple to fluid equations.


The works we have surveyed generally use the low density of the energetic particles to justify the focus on particle dynamics in a background plasma that is secondarily important.  Even in the most advanced simulation efforts, the feedback of energetic particles on the plasma is modeled in simple ways.
In order to study magnetic-field amplification around a shock, changes in the plasma turbulence induced by the energetic particles, changes in the structure of the shock, and destabilization of various instabilities around the shock,  the effects of energetic particles on the background plasma need to be included, and modeled as accurately as possible.  Because of high computational demands, many numerical studies have been performed in two dimensions where cross-field transport cannot be treated consistently.  Since two dimensional turbulence is characteristically different from three-dimensional turbulence \citep{kraichnan1971inertial,pouquet1978two}, studying the interaction between turbulence and energetic particles requires a three-dimensional code.

\section{The benefits of PDF closures \label{sec:closures}}




A successful model for plasma interaction with energetic particles requires detailed attention to closure theory and physical insight into the problem at hand.  When results from a closure model are
measured against experimental data or against results from different types of simulations, and are found to be physically correct, elements of the underlying physical motivation are confirmed.  Closure theory adopts one of two approaches: (1) to solve for approximate statistics of an exact equation, or (2) to solve for exact statistics of approximate model equations.  The first tactic is typically expressed in cumulant-discarding closures \citep[see, e.g.][]{krommes2002fundamental}, including quasinormal approximations, which yield continuum equations for the velocity or momentum of a plasma.   For example, in a cumulant-discarding closure, a number of coupled equations are produced for the transport of a quantity $Y_i$ for plasma species $i$.  This quantity could be, for example, the mass fraction of each species present, the velocity, or the specific internal energy. 
The resulting equations have the form:
\begin{eqnarray} \label{generalcumeq}
\frac{\partial \rho Y_i}{\partial t} + \nabla \cdot \left( \rho u Y_i \right) - \nabla \cdot \left( \rho D_i \nabla Y_i \right) = S_i~.
\end{eqnarray} 
Here $u$ is a fluid velocity, $D_i$ is a diffusion coefficient, $\rho$ is a density, and $S_i$ is a source term for species $i$.  
As an example, if we are dealing with a single species, and $Y_i = 1$ is its mass fraction, we recover
\begin{eqnarray} \label{continuityeq}
\frac{\partial \rho}{\partial t} + \nabla \cdot \left( \rho u \right) = S_\rho~.
\end{eqnarray} 
This is the well-known continuity equation; a source term is included on the right hand side if mass is added or removed from the system, but for many problems this source term is zero.  This equation is not closed because it includes a velocity $u$.  We therefore seek the equation for $Y_i=u$, and we recover
\begin{eqnarray} \label{momentumeq}
\frac{\partial \rho u}{\partial t} + \nabla \cdot \left( \rho u u \right) - \nabla \cdot \left( \rho D_u \nabla u \right) = S_u~.
\end{eqnarray} 
This is reminiscent of the momentum equation from the compressible Navier-Stokes equations, but a model has been assumed for the stress tensor $\nabla \cdot \tau = \nabla \cdot ( \rho D_u \nabla u)$; this model corresponds to a Newtonian fluid.  When $D_u$ is constant -- a common assumption -- this term can be expressed as simple viscous dissipation $\nu \nabla^2 u$.  The general form of eq.~\eqref{generalcumeq} can be extended to describe any number of species which experience different kinds of diffusion, and who have mass fractions that can change according to the source terms chosen.

The major limitation of truncating a cumulant-discarding closure is that in discarding and approximating higher-order cumulants, some physical information is removed.  Eq.~\eqref{momentumeq} includes terms in $\rho u u$ which must be either modeled or determined by a higher order equation.  The solutions to the closed equations may correspond to ill-behaved or unrealistic probability density functions (PDFs).  For example, one common problem is that  cumulant-discarding closures may allow the kinetic energy, a positive quantity, to become negative; this violates ``realizability''.  The realizability of any specific cumulant-discarding closure must be established \citep[e.g.][]{turner2002eddy,frederiksen2023realizable,girimaji2024turbulence}.  
Fluid models for energetic particles have commonly used cumulant discarding closures, and in most cases have used the most severe truncation of the cumulant expansion possible.   

A closure for the PDF itself does not have this disadvantage of possibly violating realizability.   {Moreover,} PDF methods have been refined considerably over the last two decades in order to model chemically reacting turbulent flows and combustion \citep{tirunagari2016pdf}.  Combustion has in common with cosmic ray astrophysics that the {particle nature of the constituents} is important; {in addition,} chemical reactions change the energetics of both the constituents and the surrounding flow.  In combustion problems there may be tens or even hundreds of chemical species, which can have many kinds of chemical reactions.   In contrast to the equations produced by a cumulant-discarding closure, a simulation of such a PDF closure solves an equation for the one-point one-time Eulerian joint PDF for all species, $P_Y (x,u,\phi,t)$.  This type of equation typically has the form
\begin{eqnarray} \label{pdfeq1}
\frac{\partial \bar{\rho} P_Y}{\partial t}
 + \frac{\partial}{\partial x_i} \left( \bar{\rho} u_i P_Y \right) 
 + \left(\rho g_i + \frac{\partial \bar{p} }{\partial x_i} \right) \frac{\partial}{\partial u_i} P_Y
+ \frac{\partial }{\partial \phi_\alpha} \left( \bar{S_\alpha}  \bar{\rho} P_Y \right) 
- \delta_{\alpha(h)} \frac{\partial }{\partial \phi_\alpha} \left( \dot{Q}_{\mathsf{rad,em}}  P_Y \right) 
= 
\\ \nonumber
\frac{\partial}{\partial u_i} \left[ \left\langle \left(-\frac{\partial \tau_{ji} }{\partial x_j}+ \frac{\partial p'}{\partial x_i} \right) \Bigg| u, \phi \right\rangle P_Y \right] 
+ \frac{\partial }{\partial \phi_\alpha} \left[  \left\langle \frac{\partial J_i^\alpha}{\partial x_i} \Bigg|  u, \phi \right\rangle P_Y \right] 
- \delta_{\alpha(h)} \frac{\partial }{\partial \phi_\alpha} \left( \langle \dot{Q}_{\mathsf{rad,em}} | u, \phi \rangle P_Y \right) 
~.
\end{eqnarray} 
Here $u$ is a velocity and $\phi$ are the composition variables, which usually include the species mass fractions and enthalpy.  The mean and fluctuating pressure contributions are $\bar{p}$ and $p'$, respectively.  Gravity is denoted by $g$, and $\delta$ is a 3D delta function used in a standard model for enthalpy. $\tau$ denotes the viscous stress tensor, $\dot{Q}_{\mathsf{rad,em}} $ is the volume rate of heating due to
radiation emission, and $J$ is the molecular fluxes of the composition variables.  $S_\alpha$ is a source term for the composition variables described as the species chemical production rate.
 The PDF and other quantities in eq.~\eqref{pdfeq1} have been Favre averaged using the density $\rho$; this density-weighted averaging is typically used for compressible flows.  This is called the velocity-composition PDF equation \citep{pope1985pdf,haworth2010progress}, and among other approximations, it is relevant to low-Mach turbulent flows.   Eq.~\eqref{pdfeq1} is a kinetic equation.  Because there can be a large number of composition variables $\phi$, for some reactive flow problems this equation would need to be solved in a dimensional space that included hundreds of variables.  Further modeling to make such equations more efficient has thus clearly been a high priority for the reactive flow community \citep{pope1981monte,pope1994lagrangian,raman2005filtered,walia2024filtered}.
 

\section{PDF methods and sparse Lagrangian models for reactive flows \label{sec:chemmodels}}

The generalization and cross-pollination of PDF closure methods and the mapping closure \citep{pope1991mapping,gao1991mapping,chen1989probability,pan2012pollution} concept has lead to the framework called the multiple mapping conditioning (MMC) for
reactive flows \citep{cleary2011multiple,klimenko2003modeling}.   
 MMC represents roughly an order of magnitude reduction compared to other PDF models. It also represents at least two orders of magnitude reduction in computational cost compared to DNS for a reacting system \citep{vo2017multiple}.  Taking a step further, the solution of MMC models using sparse Lagrangian methods has offered a new way forward for challenging large-scale problems.  MMC generalizes the mapping closure in the sense that the mapping closures for combustion originally assumed homogeneous turbulence; the MMC can cope with inhomogeneous and anisotropic flows \citep[e.g.][]{cleary2011detailed,salehi2017sparse}.

MMC {has been used} for modelling a range of combustion regimes, as well as non-premixed, partially premixed, and premixed combustion \citep[e.g.][]{devaud2013deterministic,ge2013comparative,varna2017multiplep2,vogiatzaki2011stochastic,vogiatzaki2015mixing,wandel2013hybrid,salehi2017sparse,sundaram2017pdf,galindo2017mmc}.  There are fundamentally three key concepts that together make up the MMC approach: (1) the separation of the potentially large number of species into major and minor species, (2) the application of a mixture model, and (3) the use of a mapping closure.
The equation that is the usual starting point for the MMC is
\begin{eqnarray} \label{pdfeq2}
\frac{\partial \bar{\rho} P_Y}{\partial t}
 + \frac{\partial}{\partial x_i} \left( \bar{\rho} \bar{u_i} P_Y \right) 
 + \left(\rho g_i + \frac{\partial \bar{p}}{\partial x_i} \right) \frac{\partial}{\partial u_i} P_Y
+ \frac{\partial }{\partial y_i} \left( \bar{S_i}  \bar{\rho} P_Y \right) 
+ \frac{\partial^2 }{\partial y_i y_j} \left( \bar{D}_{ij}  \bar{\rho} P_Y \right) 
= 0~.
\end{eqnarray} 
Here the quantities $\bar{\rho}$, $\bar{U}$, $\bar{S}$, and $\bar{D}$ are conditional averages representing the density $\rho$ and velocity $u$, a source term $S$, and a diffusion coefficient $D$.  This equation can be obtained by neglecting the enthalpy in eq.~\eqref{pdfeq1}, and applying a model for diffusion.
Here $x$ is the usual spatial coordinate, and $y$ is the variable for physical quantity $Y_i$.    In the context of reactive flows, the $y$ variables track mass fractions and other quantities for the chemical reactions.



\subsection{Major and minor reacting species}

Large numbers of species are often present for combustion problems, so for the sake of efficiency, the MMC categorizes the species  as either ``major species'' or ``minor species''; this kind of different treatment is a common step in the reactive flow community.  For a total number of species $n_s$, there are $n_m$ major species denoted $Y^m$, and $n_\alpha$ minor species denoted $Y^\alpha$, so that $n_s = n_m + n_\alpha$.  Major species are denoted with Latin indices and minor species are denoted by Greek indices.  Major species are allowed to fluctuate freely; minor species are limited to fluctuate in the same way that the major species do.  The division into major and minor species simplifies the calculation of the joint PDF, which can then be replaced by the  PDF of major
species, multiplied by a conditional mean that restricts the conditional distributions of the minor species to be delta functions
\begin{eqnarray} \label{marginalpdf}
P_Y = P_{Y^m} \delta (Q_\alpha - y^{\alpha})~.
  \end{eqnarray} 
Here $P_Y$ is the joint PDF of all species, $P_{Y^m}$ is the joint PDF of major species, and $Q_\alpha = \langle Y_\alpha | Y^m = y^m \rangle$.  Using this decomposition in a transport equation like \eqref{pdfeq2} produces two independent equations for $P_{Y^m} $ and $Q_\alpha$
  \begin{eqnarray} \label{evolutionmarginalpdf}
\frac{\partial \bar{\rho} P_{Y^m}}{\partial t}
 &+& \nabla \cdot \left( \bar{\rho} \bar{U} P_{Y^m} \right) 
+ \frac{\partial }{\partial y_i} \left( \bar{S_i}  \bar{\rho} P_{Y^m} \right) 
+ \frac{\partial^2 }{\partial y_i y_j} \left( \bar{D}_{ij}  \bar{\rho} P_{Y^m} \right) 
= 0~,
\\ \label{deltaeveq}
\frac{\partial Q_\alpha}{\partial t}
 &+& \bar{U} \nabla \cdot Q_\alpha
+  \bar{S_i} \frac{\partial }{\partial y_i} Q_\alpha
+ \bar{D}_{ij}  \frac{\partial^2 }{\partial y_i y_j} Q_\alpha 
= 0~.
  \end{eqnarray} 
This amounts to the use of a conditional moment closure (CMC) model \citep{klimenko1999conditional} for the minor species, who experience ``surrogate mixing''.
  With the split into major and minor species, the reference space that the mapping closure makes use of  only requires $m$ reference variables.
This different treatment of major and minor species models the direct influence of mixing on the fluctuations of the major species.
Any secondary influences of the minor fluctuations on the
major fluctuations through chemical source terms or other
nonlinear mechanisms is neglected.
These equations are not yet closed because they require the input of drift $\bar{U}$,  diffusion $\bar{D}_{ij}$, and source terms $ \bar{S_i}$ that represent mixing and reactions.



 \subsection{Mapping closures and Mixture Models}
     
The model requires additional closures for the mixing processes \citep{pope1991mapping}.
Several variations of these closures exist; however the key feature common to all MMC models is the use of reference
variables which are related to the physical quantities.
MMC and mapping closures make use of a reference space that has a PDF that is prescribed a priori, and can be simple, e.g. Gaussian. 
A set of mapping functions is calculated that has as input reference space variables $\xi$, and has as output the range of expected values
of all the species $Y$.  This amounts to solving for a set of mapping functions $X(\xi;x, t) = (X_1, . . . ,X_m)$ such that $X$ is
statistically equivalent to the actual variables $Y$.  All moments of $Y$ can then be calculated from the PDFs of $\xi$ and $X(\xi)$.
If needed, additional kinetic equations like eq.~\eqref{pdfeq2} can be used to evolve the distributions of the reference variables as described by \citet{varna2017multiplep1}.

The closures for the mixing processes have two components; the coefficients for the major species are calculated using the reference space, and for the minor species they are calculated through a further mixing operation.  The major species mixing processes are local in the space of the reference
variables $\xi$.  At each time step all particles in the domain are
grouped in pairs (without replacement). Particle pairs are assigned by taking the minimum of a normalized square distance, 
which includes contributions both in real space $x$ and in the reference variable space $\xi$.   Once selected, the pair of mixing particles alters their mass fractions and enthalpies linearly and discretely over a time step.  The mixing models which are commonly used for PDF modeling are Curls \citep{ren2004investigation} and modified Curls models, and the interaction by exchange with the mean (IEM) model \citep[e.g.][]{wandel2024numerically}.
Equations that result from mapping closures have been demonstrated to capture both qualitative and quantitative features of intermittent turbulence in the form of highly non-Gaussian PDFs.  
  


\subsection{Simulation strategy for the sparse-Lagrangian MMC}

Simulations where the number of Lagrangian particles is much smaller than the number of Eulerian grid points are described as sparse Lagrangian simulations, as opposed to intensive Lagrangian simulations. The MMC approach has been developed in conjunction with sparse Lagrangian simulation methods. This approach, sometimes called MMC-LES or MMC-DNS, replaces the simple Gaussian reference variables by variables generated by Lagrangian tracer particles in LES or DNS simulations.  This provides more realistic values for first conditional moments.

The sparse-Lagrangian MMC model consists of two separate solvers: the fluid
LES or DNS solver and the stochastic Lagrangian particle solver.  Two-way coupling is implemented between the Eulerian LES fields and Lagrangian particles.
The PDF transport equations of the form in eq.~\eqref{marginalpdf} are
replaced by the equivalent Lagrangian transport equations \citep{zwanzig2001nonequilibrium,pope1985pdf,liu2024sparse,neuber2019sparse,huang2020application}:
\begin{eqnarray}\label{lagtrans1}
d x_i^q &=& \left[u_i + \frac{1}{\bar{\rho}} \frac{\partial}{\partial x_i} \bar{\rho} D_{\mathrm{eff}} \right]^q dt + (\sqrt{2 D_{\mathrm{eff}} })^q d w_i~,
\\ \label{lagtrans2}
d Y_m^q &=& (W_m^q + S_m^q) dt ~,
\\ \label{lagtrans3}
d h &=& \left[ W_h^q + S_h^q + \left(\frac{1}{\bar{p}}\frac{D \bar{p}}{D t} \right)^q +  \left(\frac{1}{\bar{p}} \sigma_{ij}  \frac{\partial u_i}{\partial x_j} \right)^q  \right] dt ~.
\end{eqnarray}
Here $q$ is a particle index, $w_i$ is a Wiener process, $\sigma_{ij}$  is the compressible pressure work, $W_m^q$ is a chemical source term, and $W_h^q$ is the heat loss. $S_m^q$ is a mixing term supplied by the mixture model for mass fractions, and $S_h^q$ is a mixing term for the enthalpy.  $D_{\mathrm{eff}}$ is an effective diffusion coefficient that includes molecular diffusion, turbulent diffusion, momentum diffusion, as well as other contributions. These are It\^o stochastic differential equations.  Having produced physically relevant fields and initialized Lagrangian tracer particles, the simulation
proceeds with the following algorithm:
\begin{enumerate}
\item The solution for the next time step is obtained using fluid equations in the LES or DNS.
\item Fluid values for velocity, pressure, mixture fraction, and material derivatives are interpolated to particle positions.
\item The Lagrangian transport equations, eqs.~\eqref{lagtrans1}--\eqref{lagtrans3} above are solved for the position of the particles at the next time step.
\item Density, mass fractions, and enthalpy are passed from the Lagrangian solution to the LES or DNS by integrating over the particles \citep{cleary2011detailed,muradoglu2001hybrid}. 
\end{enumerate}
The details of this final step are worth making more explicit.  The calculation for density is \citep{raman2007consistent}
\begin{eqnarray}
\frac{1}{\bar{\rho}} = \sum_{i=1}^N \frac{w_i}{\rho(\phi_i)}~.
\end{eqnarray}
Namely, using the particle's mass fractions $\phi_i$ and the appropriate weights $w_i$, the density in any control volume is calculated from the N particles in that control volume.  An Eulerian PDE for the enthalpy is solved using input from the particles.  Further transport equations for the equivalent
compositions are also solved \citep[][cf. eq. (12)]{breda2023coupling}.

The sparse-Lagrangian MMC method outlined here has been demonstrated to provide a good approximation to the large-scale structures of
reactive scalars.  It also provides a weak approximation for the overall distributions of the composition
scalars. This is a general attribute of sparse-Lagrangian simulation methods.





\section{A new model for cosmic ray interaction with a background plasma \label{sec:crmodels}}

A particularly desirable advantage of the sparse-Lagrangian MMC framework for reactive flows is that {it has} been developed to be efficiently solved for large-scale problems.  For astrophysical plasmas, this framework can be coupled to a LES simulation to treat the extremely {large scales} relevant to the heliosphere, supernovae, galactic winds, or the galactic dynamo.

As we have seen in Section~\ref{sec:chemmodels}, the sparse-Lagrangian MMC framework for reactive flows has some fundamental similarities to the hybrid approach for cosmic rays that uses fluid models for the background and a Fokker--Planck approach for the particles.
Here is a Fokker--Planck equation formulated to study cosmic rays, using the notation of \citet{schlickeiser2013cosmic}, eq. (12.1.26)
\begin{eqnarray} \label{fokkerplanckeq1}
\frac{\partial F_\alpha}{\partial t}
 + \frac{\mu p}{\gamma m_\alpha} \frac{\partial }{\partial Z}   F_\alpha
- \epsilon \Omega \frac{\partial }{\partial \phi}  F_\alpha
= S_\alpha (x, p, t) + \frac{1}{p^2} \frac{\partial }{\partial x_\sigma} \left(p^2 D_{\sigma,  \nu}  \frac{\partial }{\partial x_\nu} F_\alpha \right) ~.
\end{eqnarray} 
In this equation $F_\alpha$ is the PDF,  $\Omega$ is the gyrofrequency of a charged particle, $\epsilon$ is the sign of the charge of the particle, and $\gamma$ is the Lorentz factor.  Considering spherical coordinates in momentum space where $\theta$ is one angular coordinate and $\phi$ is another, the pitch angle cosine is $\mu = \cos \theta $; the {6-dimensional} coordinate is then $x_\sigma = (p, \mu, \phi, X,Y,Z)$.  
{Here is another} Fokker--Planck equation formulated to study cosmic rays, using the notation of \citet{zweibel2017basis}, eq. (8)
\begin{eqnarray} \label{fokkerplanckeq2}
\frac{\partial f}{\partial t}
 + v \cdot \nabla f 
=  \frac{\partial }{\partial \mu} \left( D_{\mu p} \frac{\partial f}{\partial p} + D_{\mu \mu} \frac{\partial f}{\partial \mu} \right)
+ \frac{1}{p^2}\frac{\partial }{\partial p}  p^2 \left( D_{p p} \frac{\partial f}{\partial p} + D_{p \mu} \frac{\partial f}{\partial \mu} \right)
~.
\end{eqnarray} 
Here $f$ is the PDF, where other variables have the same meaning as in eq.~\eqref{fokkerplanckeq1}.  Both of these Fokker--Planck equations operate in {7-dimensional} space, with 3 spatial dimensions, 3 momentum dimensions, and time.  Still for cosmic ray Fokker--Planck equations, the challenge is capturing the relevant physics for diffusion coefficient $D$ with sufficient fidelity.

There are several differences between eq.~\eqref{pdfeq1} and either of these Fokker--Planck equations formulated for cosmic rays.  One difference is that eq.~\eqref{pdfeq1} has been generalized to include a source term that results from mixing, and thus has the potential for chemical reactions to move energy from one species to another.  The energization of particles is a situation where length and time scales below the size of an achievable Eulerian grid are expected to contribute to the physical process; this is a problem that particle acceleration in plasmas holds in common with chemical reactions.
For the study of the acceleration of cosmic rays, an astrophysical plasma MMC framework will have a source term; particles should be able to be energized so that cosmic rays are created.   The framework of reactive flows allows for the classification of different chemical species, and the re-classification of those species through chemical reactions.  The analogous concept for the problem of cosmic rays is the idea of energetic species; to our knowledge, this is a new idea that has not before been articulated for the modeling of cosmic rays.   

\begin{figure}
\begin{center}
\resizebox{4in}{!}{\includegraphics{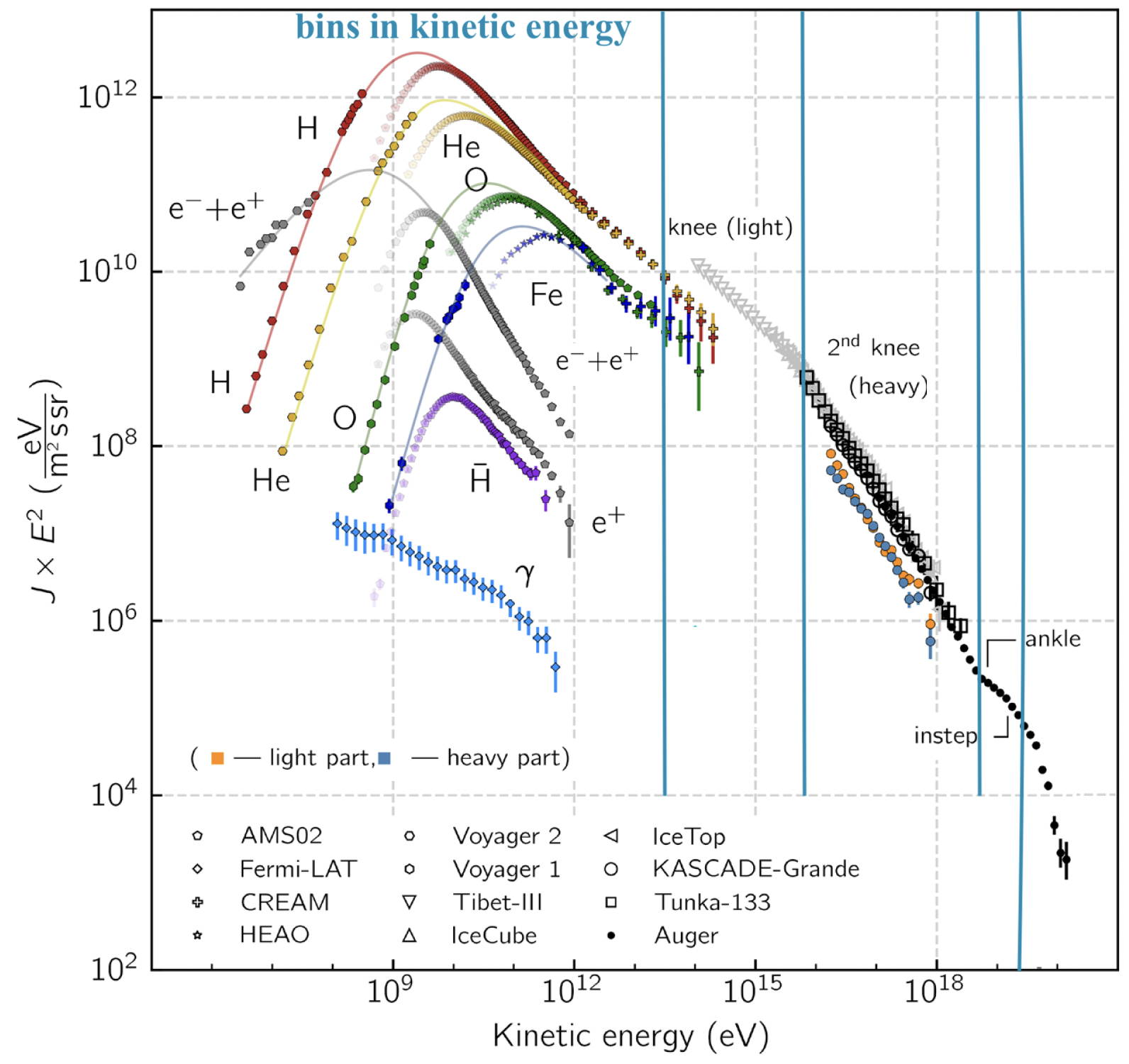}}
\caption{Possible bins in kinetic energy that could define energetic species, drawn on the cosmic ray spectrum, figure from the review of \citet{ruszkowski2023cosmic}, original measurements from a variety of instruments \citet{lenok2021measurement,stone2019cosmic}.
\label{fig:spectrum}}
\end{center}
\end{figure}

The cosmic ray spectrum is famous for having named regions: the knee, the second knee, the ankle, the instep, as well as a bulk population of lower energy particles.  These different energetic regions are the natural energetic species to define for a sparse-Lagrangian MMC model for cosmic rays (see for example the bins drawn in Figure~\ref{fig:spectrum}).  The MMC framework does not require the mass fractions of species to be large or roughly equivalent.
Using a MMC model designed for energetic reactions provides a natural mechanism for an ordinary plasma particle to be transformed into a knee particle, or an instep particle to be transformed into an ankle particle.  Using the major-minor {breakdown} of species, some minor energetic species can be constrained to fluctuate similarly to the background plasma, while other minor energetic species can fluctuate similarly to the highest energy cosmic rays.  The most attractive aspect of creating different energetic species is that the two-way interaction of cosmic ray particles on background plasma particles can be treated with greater fidelity than can be included in a hybrid fluid/Fokker--Planck solver.  This concept will need further exploration through simulation data so that the number of major and minor energetic species needed to describe the {particle acceleration process} can be established. 

The astrophysical plasma MMC framework will also take advantage of the MMC's ability to calculate diffusion coefficients on the fly from simulation data.  Exploiting this ability allows the diffusion coefficient to be dependent on the exact local magnetic field.  In the presence of magnetic fields, diffusion and mixing processes can become anisotropic on both large and small scales, and can be affected by Alfv\'enic fluctuations, a fundamental difference from neutral-fluid diffusion \citep{homann2007lagrangian, busse2007statistics, busse2008diffusion, busse2010lagrangian, pratt2017extremehull,pratt2020lagrangian,pratt2022reynolds}.  Moreover, the use of diffusion coefficients calculated from simulation data will help to resolve the question of what kind of non-Fickian diffusion cosmic rays experience (with reference to our discussion of telegraph equations and hyperdiffusion in Section~\ref{sec:existingmodels}).
  The use of sparse Lagrangian methods for the astrophysical plasma MMC framework  will make it possible to achieve more accurate mixing and diffusion than comparatively simple Fokker--Planck solvers.  Using different energetic species in {an} astrophysical plasma MMC framework provides the ability to make the advection and diffusion coefficients specific to the energetic species as well.



\section{Summary}

In their seminal paper, \citet{klimenko2003modeling} described the MMC
approach as   ``more akin to a certain framework that can be used to formulate various specific models.''  In this work we have reviewed the extensive literature on this framework, as well as the solution methods devised for  the problem of reactive flows.  

We have outlined how an approach based on the sparse-Lagrangian MMC method for astrophysical plasmas and cosmic rays would improve on current methods that employ hybrid fluid/Fokker--Planck solvers.  We have pointed out several technical aspects of this model that are expected to lead to higher flexibility and higher fidelity simulations in general.  These include: better modeling of two-way interactions between cosmic rays and a background plasma, better diffusion models, and the ability to perform simulations at the largest spatial scales.    We have developed these ideas in order to  simulate cosmic ray acceleration as well as its feedback on turbulent astrophysical plasmas.

\section*{Acknowledgements \label{sec:ack}}
{\small  This work was supported by the National Science Foundation under grant no. PHY-1907876.
Part of this work was performed under the auspices of the U.S. Department of Energy by Lawrence Livermore National Laboratory under Contract DE-AC52-07NA27344.  LLNL-TR-2014987.
}

\bibliographystyle{unsrtnat}
{\small \bibliography{pitch}}



\end{document}